\documentclass[aps,prd,notitlepage,nofootinbib,onecolumn,groupedaddress,showkeys,floatfix]{revtex4-2}
\usepackage{graphicx}
\usepackage{amssymb}
\usepackage{amsmath}
\usepackage{bm}

\begin{document}

%\preprint{}

\title{Back to Bohr: Quantum Jumps in Schr\"odinger's Wave Mechanics}

% Please use the appropriate macro for each type of information


%%\author{Student 1}
%%\email{student.1@someuni.somecountry}
%%\affiliation{Department of Physics and Engineering Physics, 
%%University of Saskatchewan, 116 Science Place, Saskatoon, Canada SK S7N 5E2}
%%\affiliation{Department of Physics, SomeUniversity, SomeCountry}

%\author{Co Author}
%\email[]{rainer.dick@usask.ca}
%\email{@}

\author{Rainer Dick}
%\email[]{rainer.dick@usask.ca}
\email{rainer.dick@usask.ca}
%\thanks{}
\affiliation{Department of Physics and Engineering Physics, 
University of Saskatchewan, 116 Science Place, Saskatoon, Canada SK S7N 5E2}


%\date{\today}

\begin{abstract}
  The measurement problem of quantum mechanics concerns the question under
  which circumstances coherent wave evolution becomes disrupted to produce
  eigenstates of observables, instead of evolving superpositions of eigenstates.
  The problem needs to be addressed already within wave mechanics, before
  second quantization, because low-energy interactions can be dominated by
  particle-preserving potential interactions.

  We discuss a scattering
  array of harmonic oscillators which can detect particles penetrating the
  array through
  interaction with a short-range potential. Evolution of the wave function
  of scattered particles, combined with Heisenberg's assertion that
  quantum jumps persist in wave mechanics, indicates that the wave function will
  collapse around single oscillator sites if the scattering is inelastic, while it
  will not collapse around single sites for elastic scattering.

  The Born rule for position observation
  is then equivalent to the statement that the wave function for
  inelastic scattering amounts to an epistemic superposition of possible
  scattering states, in the sense
  that it describes  a sum of probability amplitudes for inelastic scattering off different
  scattering centers, whereas at most one inelastic scattering event can happen at any
  moment in time. 

  Within this epistemic interpretation of the wave function, the actual underlying
  inelastic scattering event corresponds to a quantum jump, whereas the continuously
  evolving wave function only describes the continuous evolution of probability
  amplitudes for scattering off different sites.
  Quantum jumps then yield definite position
  observations as defined by the spatial resolution of the oscillator array.
\end{abstract}

% insert suggested PACS numbers in braces on next line %%separate by commas
%\pacs{ , }  
% insert suggested keywords - APS authors don't need to do this%%separate by commas

\maketitle

%\section{Introduction}
%\subsection{}
%\subsubsection{}

%%%%%%%%%%%%%%%%%%%%%%%%%%%%%%%%%%%%%%%%%%%%%%%%%%%%%%%%%%%%%%%%%%%%%%%%%%%%%%%%
%%%%%%%%%%%%%%%%%%%%%%%%%%%%%%%%%%%%%%%%%%%%%%%%%%%%%%%%%%%%%%%%%%%%%%%%%%%%%%%%
\section{Introduction}\label{sec:intro}

The problem of definite measurement outcomes from observations of quantum systems
which initially exist in a superposition of many possible eigenstates of the observed variable,
is best illustrated and discussed in terms of basic examples. Unfortunately, this comes at
the price of some redundancy with previous discussions, and at the risk of annoying knowledgeable
readers. My excuses are twofold: On the one hand, the measurement problem
is too important from a principal perspective to only be left
to the initiated. On the other hand, staying close to actual manifestations of the
measurement problem improves clarity and efficiency of the discussion.
Too much abstraction, on the other hand, can easily lead to elaborate proposals
which instantly fall apart when gauged against actual examples of observations.

Position observations for electrons provide particularly
neat illustrations of the measurement problem.
 For example, in the single-electron
diffraction observations by Bach \textit{et al.}, single electrons from a 2 micrometre 
wide collimated beam diffract off two slits (each 62 nm wide and 272 nm apart),
to create a more than 4 mm wide well-resolved two-slit interference pattern at the location
of a microchannel plate (MCP) detector, after accumulation of many single-electron signals
\cite{rf:bach}. The authors give $238.2\pm 6.6$ micrometres
for the position resolution of the imaging system. The standard interpretation of quantum
mechanics of this experiment, to which we adhere, is that the single-electron
position probability density $|\psi(\bm{x},t)|^2$, at the location of the detector, have a lateral
width of more than 4 mm, with seven clearly identifiable peaks,
after traversing the double-slit and travelling all the way to the detector. However, when an
electron decribed by that wave function hits the detector, it creates a particle-like signal
in the sense that the signal is confined to within the spatial resolution of the detector,
which is much smaller than the inherent width of the wave function.
This is exactly how we define a pointlike particle signal \cite{rf:rdjgps}:
The spatial width of the signal
is determined by detector resolution, but not by the width of the incoming
wave function.
Stated differently, impact of the electron on the MCP creates a pointer state that
corresponds to excitation of one channel (or at most a few channels within the spatial
resolution of the imaging system), but not to excitation of several
distinct macroscopically
separated channels. The latter outcome would sometimes also be denoted as a superposition
of pointer states, which would correspond to simultaneous detection of several positions
of the electron.

The measurement problem for position observation then concerns the following question:
The electron wave function coherently passes through the double-slit to evolve
into the macroscopically wide wave of a double-slit interference pattern, and yet, when that
wave hits the detector, it triggers a pointlike signal with a much smaller width.
Why does the wave not light up an interference pattern across the MCP upon every impact?
And how come that accumulation of many consecutive pointlike
impacts creates that very interference pattern?

At this point we might just sit back and relax
and remind ourselves of the Born interpretation of the wave function. This is exactly
what the wave function is supposed to do: ``If we ask the electron where it is, it
responds with a pointlike signal with a probability density distribution given
by $|\psi(\bm{x},t)|^2$.'' Alas, it is not that simple. Both the passage through
the double slit and the registration in the MCP detector are ultimately scattering
events. How would the electron (or its wave function) ``know'' that scattering
off the double-slit is not a detection event, and therefore scattering occurs coherently
off both slits in a wavelike manner,
but the scattering in the MCP is a detection event and therefore scattering and subsequent
trigger of an electron avalanche occurs in one channel in a particle-like manner,
instead of coherently exciting many
channels over a 4 mm wide area?

The question is even more striking if we replace the double slit with elastic scattering off
a metal surface, as in a low-energy electron diffraction (LEED) event.
Electron scattering off the metal sample produces a macroscopic wave function that
can (upon accumulation of many single-electron signals) create a macroscopic interference
pattern once the electrons scatter in a fluorescent screen.
Again, the electron wave function after scattering off the metal surface is a coherent
superposition of scattering off different scattering centers, but the final signal
is produced by scattering off only a single scattering center in the fluorescent
screen\footnote{We were focusing on the double-slit example, because it is so
well-documented as a single-electron
diffraction experiment. To my knowledge, LEED has not be done as a deliberate
  single-electron diffraction experiment, although this would be very nice, too.
  Of course, one would again expect to see gradual accumulation of the interference
  pattern from coherent elastic scattering off a metal surface, but the demonstration
  of initial elastic wave scattering followed by inelastic
  particle scattering would be even more striking.}.

What the two different kinds of scattering events separates in either case is
that the first scattering event involves elastic wavelike scattering,
which generates the interference pattern, while the second particle-like scattering
event is inelastic and generates the individual position signal. Energy transfer between
the particle and the detector apparently breaks the coherent smooth
evolution of the wave function.

It has been shown before that inelastic potential scattering off the short-range potential
of localized scattering centers leads to a contraction of the scattered
wave function \cite{rf:rd2024a}.
However, that by itself does not necessarily imply
a breakdown of smooth evolution of the wave function during inelastic scattering.
On the other hand, the breakdown of smooth evolution is required to explain the absence of
superposition effects from
inelastic scattering during the position observation: The many-peaked probability density
$|\psi(\bm{x},t)|^2$ from interference effects after the initial elastic scattering does not
generate more than one position signal at a time upon inelastic scattering in the detector.
The scattered wave function contracts only here or there, but it never contracts both
here and there.
This absence of superposition effects after inelastic scattering is a clear indication of the
disruption 
of smooth wave function evolution during energy exchange between one quantum system and another
quantum system, or equivalently, between two components of a quantum system.

A break in smooth evolution of a system due to energy exchange is a quantum jump, and it can be
argued in the framework of second quantization that the second-quantized time-dependent
Schr\"odinger equation does not describe the smooth evolution of quantum states, but instead describes
the smooth evolution of scattering matrix elements as probability amplitudes for quantum
jumps between different sectors in Fock space \cite{rf:rdshpmp,rf:rdjgps}.

However, this observation is not sufficient to resolve the measurement problem, because
low-energy interactions, including low-energy interactions between particles and detectors, are
dominated by particle-preserving potential interactions and do not inherently require second
quantization nor transitions between different sectors in Fock space\footnote{For example,
the ratio of the scattering matrix elements for electron scattering $\bm{p}\to\bm{p}'$
off atoms or other electrons,
from photon exchange versus Coulomb scattering
in Coulomb gauge is $|S_{fi}^{(\gamma)}/S_{fi}^{(C)}|\lesssim |\bm{p}+\bm{p}'|^2/4m^2c^2$
in the low-energy limit. See, e.g.~Secs.~23.2 (especially Eq.~(23.52)) and 23.4 in
Ref.~\cite{rf:rdqm3ed}.}.
Transitions between sectors
of Fock space are helpful as clear indicators of quantum jumps, but we need
quantum jumps also within the confines of Schr\"odinger's wave mechanics to resolve the
measurement problem.
Indeed, within the framework of wave mechanics, Heisenberg had asked soon after
inception of the Schr\"odinger equation whether it only disguises quantum jumps through a smooth
statistical evolution \cite{rf:wh1}. He repeated that suspicion in his lectures
at the University of Chicago, see Appendix 5 in \cite{rf:wh2}. Heisenberg's suspicion leads
to a tempting resolution of the measurement problem: Every observation necessitates
energy transfer between detector and observed system, and if this energy transfer disrupts
smooth wavelike evolution through a quantum jump, the annoying problem
of superposition of different pointer states as a consequence of coherent Schr\"odinger
evolution disappears. 

To explain this, we introduce an oscillator array as a model system for
low-energy particle scattering in Sec.~\ref{sec:array}.
The oscillator array serves both as a particle diffractor
through elastic scattering, and as a particle detector through inelastic scattering.
Interpreting the wave function after inelastic scattering as an ontic quantum state
would inevitably predict superposition of position signals from different locations,
in contradiction to observation.
On the other hand, we know that the wave function
from a single \textit{elastic} scattering event, as a coherent superposition of elastic scattering
off different scattering centers, should have an ontic interpretation, because
accumulation of position signals through single-electron inelastic scattering reveals
the interference pattern predicted by coherent scattering off different centers.
The difference in the scattering events is energy transfer, and we can break the
dichotomy in quantum evolution between the different scattering events if we postulate
that energy transfer evolves discontinuously through quantum jumps. Wavelike behavior
then emerges from elastic scattering, whereas particle-like behavior emerges from
inelastic scattering.

Our conclusions are summarized in Sec.~\ref{sec:conc}.
The apparent discrepancy of observation and coherent wavelike evolution of wave functions
has led to epistemic interpretations of the wave function as a tool to predict possible
outcomes of quantum evolution for
given initial conditions \cite{rf:epi1,rf:epi2,rf:epi3,rf:epi4}. Our results suggest
that wave functions can describe ontic evolution of quantum states as long as energy transfer
within multi-partite systems can be neglected, while they describe epistemic superpositions
of possible ontic outcomes as soon as energy transfer becomes relevant.

\section{Particle diffraction and detection}\label{sec:array}

We model particle diffraction and detection through interaction of a particle
(with coordinates $\bm{x}$ and momentum $\bm{p}$) with an array of $N$
scattering centers at positions $\bm{a}_I$,
\begin{eqnarray}\nonumber
  H&=&H_0+V
  \\ \label{eq:model1}
  &=&\frac{\bm{p}^2}{2m}+\sum_{I=1}^N\left(\frac{\bm{P}_I^2}{2M}+U(\bm{y}_I-\bm{a}_I)\right)
   +\sum_{I=1}^N V(\bm{x}-\bm{y}_I),
\end{eqnarray}
The potentials $U(\bm{y}_I-\bm{a}_I)$ set the internal energy levels 
of the scattering centers
while the particle is scattered through the potential $V=\sum_{I=1}^N V(\bm{x}-\bm{y}_I)$.

The Schr\"odinger equation predicts evolution of the corresponding $(N+1)$-particle
wave function $\langle\bm{x},\bm{y}_1,\ldots,\bm{y}_N|\Psi(t)\rangle$
according to
\begin{eqnarray}\label{eq:Psifi2}
  \langle\bm{x},\bm{y}_1,\ldots,\bm{y}_N|\Psi(t)\rangle&=&
  \langle\bm{x},\bm{y}_1,\ldots,\bm{y}_N|\exp[-\,\mathrm{i}H(t-t')/\hbar]|\Psi(t')\rangle.
\end{eqnarray}

We assume that the scattering centers are harmonic oscillators with equilibrium
positions $\bm{a}_I$,
\begin{equation}
U(\bm{y}_I-\bm{a}_I)=\frac{1}{2}M\Omega^2(\bm{y}_I-\bm{a}_I)^2.
\end{equation}
The detector therefore consists of $N$ ``elementary detectors'' or ``detection units'', each of which
can record the position of the incoming particle through a change in energy. The question is:
Why would not several of those detection units trigger simultaneously over the full width
of the incoming particle wave function, although the detector-plus-particle wave function
$\langle\bm{x},\bm{y}_1,\ldots,\bm{y}_N|\Psi(t)\rangle$ contains a superposition of several
excited detector units already in first order of the interaction potential $V$?

We assume that the detector was in its ground state before scattering of the
incoming particle,
\begin{eqnarray}\label{eq:Psifi0}
  \langle\bm{x},\bm{y}_1,\ldots,\bm{y}_N|\Psi(t')\rangle&=&
  \langle\bm{x}|\psi_i(t')\rangle\exp(-\,\mathrm{i}NE_{\bm{0}}t'/\hbar)
  \prod_{I=1}^N\phi_{\bm{0}}(\bm{y}_I-\bm{a}_I),
\end{eqnarray}
where $\phi_{\bm{0}}(\bm{y}_I-\bm{a}_I)$ is the ground state wave function of the
oscillator centered at $\bm{a}_I$.

Projection of (\ref{eq:Psifi2}) into the single-particle sector for the
scattered particle yields in leading order of $V$ (and after expressing
initial and final states at fiducial time $t_0=0$, as usual)
\begin{eqnarray}\nonumber
  |\psi_f\rangle&=&|\psi_i\rangle-\sum_{I=1}^N\sum_{\bm{n}\ge 0}\frac{\mathrm{i}}{\hbar}
  \int_{t'}^t\!d\tau\,\exp(\mathrm{i}\omega_{\bm{n},\bm{0}}\tau)
  \int\!d^3\bm{y}\,
  \exp\!\left(\mathrm{i}\frac{\bm{p}^2\tau}{2m\hbar}\right) V(\bm{x}-\bm{y})
  \\  \label{eq:Sfi1}
  &&\times
  \phi^+_{\bm{n}}(\bm{y}-\bm{a}_I)\phi_{\bm{0}}(\bm{y}-\bm{a}_I)
  \exp\!\left(-\,\mathrm{i}\frac{\bm{p}^2\tau}{2m\hbar}\right)|\psi_i\rangle.
\end{eqnarray}
Here, $\bm{p}$ and $\bm{x}$ are operators for the scattered particle.
The transition frequency $\omega_{\bm{n},\bm{0}}$ corresponds to the energy transfer from
the particle to the scattering center,
\begin{equation}
  \omega_{\bm{n},\bm{0}}=(E_{\bm{n}}-E_{\bm{0}})/\hbar.
\end{equation}

Projection of Eq.~(\ref{eq:Sfi1}) into position representation for the scattered particle
yields for scattered states ($\psi_f(\bm{x})\neq\psi_i(\bm{x})$)
  \begin{eqnarray}\nonumber
    \psi_f(\bm{x})&=&\frac{1}{(2\pi)^3\mathrm{i}\hbar}\sum_{I=1}^N\sum_{\bm{n}\ge 0}
    \int\!d^3\bm{y}\int\!d^3\bm{z}
  \int\!d^3\bm{x}'\int\!d^3\bm{k}\int_0^\infty\!dk'\,k'\,\exp[\mathrm{i}\bm{k}\cdot(\bm{x}-\bm{z})]
  \\ \nonumber
  &&\times
  \frac{\sin\!\left(k'|\bm{z}-\bm{x}'|\right)}{
    \pi|\bm{z}-\bm{x}'|}\, V(\bm{z}-\bm{y})
  \phi^+_{\bm{n}}(\bm{y}-\bm{a}_I)\phi_{\bm{0}}(\bm{y}-\bm{a}_I)\psi_i(\bm{x}')
  \\ \label{eq:psif1tt} %\nonumber
  &&\times
  \frac{\sin\!\left([2m\omega_{\bm{n},\bm{0}}+\hbar(k^2-k'^2)](t-t')/4m\right)}{
    \pi[2m\omega_{\bm{n},\bm{0}}+\hbar(k^2-k'^2)]/2m}\,
  \exp\!\left(\mathrm{i}[2m\omega_{\bm{n},\bm{0}}+\hbar(k^2-k'^2)](t+t')/4m\right)\!.
  \end{eqnarray}

  Eq.~(\ref{eq:psif1tt}) expresses the fact that the wave function after a single scattering event
  comprises a coherent superposition of contributions from each scattering center, 
  as confirmed for elastic scattering through the observation of interference effects.
  
  The factor in the last line of Eq.~(\ref{eq:psif1tt})
  approaches $\delta\!\left([2m\omega_{\bm{n},\bm{0}}+\hbar(k^2-k'^2)]/2m\right)$ if
  \begin{equation}
    t-t'\gg 2m/[2m\omega_{\bm{n},\bm{0}}+\hbar(k^2-k'^2)].
  \end{equation}
  For inelastic scattering, $\bm{n}\neq\bm{0}$, this condition is fulfilled for all times
  that satisfy $t-t'\gg 1/\Omega$, i.e.~in this case we find for the inelastic scattering
  contribution
  \begin{eqnarray}\nonumber
    \psi_f^{(\mathrm{inel.})}(\bm{x})&=&\frac{m}{(2\pi)^3\mathrm{i}\hbar^2}\sum_{I=1}^N\sum_{\bm{n}\neq\bm{0}}
    \int\!d^3\bm{y}\int\!d^3\bm{z}
  \int\!d^3\bm{x}'\int\!d^3\bm{k}\,\exp[\mathrm{i}\bm{k}\cdot(\bm{x}-\bm{z})]
  \\ \nonumber
  &&\times
  \frac{\sin\!\left(\sqrt{k^2+(2m\omega_{\bm{n},\bm{0}}/\hbar)}|\bm{z}-\bm{x}'|\right)}{
    \pi|\bm{z}-\bm{x}'|}\, V(\bm{z}-\bm{y})
  \\ \label{eq:psifinel}
  &&\times\phi^+_{\bm{n}}(\bm{y}-\bm{a}_I)\phi_{\bm{0}}(\bm{y}-\bm{a}_I)\psi_i(\bm{x}').
  \end{eqnarray}
  Indeed, the emergence of the energy-preserving $\delta$-function follows both from the
  sinc function in the last line of Eq.~(\ref{eq:psif1tt}), and also from the fact that fast
  oscillation of the exponential factor in the last line would also erase the scattered
  wave function unless energy is conserved.

  Since $m=511\,\mathrm{keV}/c^2$ and $\omega_{\bm{n},\bm{0}}\ge\Omega$, the normalized sinc function
  in the second line has at most {\AA}ngstr\"om-scale width if $\hbar\Omega>1\,$eV.
  Furthermore, we also assume that the scattering potential $V$ has short range.
  Approximating those two short-range factors by $\delta$-functions then yields
   \begin{eqnarray}\label{eq:psifinel2}
     \psi_f^{(\mathrm{inel.})}(\bm{x})&\propto&\sum_{I=1}^N\sum_{\bm{n}\neq\bm{0}}
     \phi^+_{\bm{n}}(\bm{x}-\bm{a}_I)\phi_{\bm{0}}(\bm{x}-\bm{a}_I)\psi_i(\bm{x}),
   \end{eqnarray}
   i.e.~the inelastically scattered piece of the wave function
   consists of $N$ pieces which are cut out from the
   incoming wave function. These $N$ pieces are centered around the detector units
   at locations $\bm{a}_I$, and the widths of those pieces
   are determined by the sizes of the excited detector states, which are
   of order $\sqrt{\hbar|\bm{n}|/M\Omega}$.
   
   If we now argue with Heisenberg that wave function evolution is only a smooth probabilistic
   approximation to quantum jumps in the detector units, then we cannot assume continuation
   of wavelike superposition at this stage, and it appears reasonable to assume that only
   one quantum jump related to an energy transfer actually happens: The inelastically scattered
   part of the wave function collapses around a single detector site, thus
   yielding only a single position signal.
   The relative magnitude of the $N$ terms in (\ref{eq:psifinel2}) determines the
   relative probability amplitudes, and the detector unit and energy transition with
   the largest overlap factor
   \begin{equation}
     P_{iI\bm{n}}=\int d^3\bm{x}\,
     |\phi^+_{\bm{n}}(\bm{x}-\bm{a}_I)\phi_{\bm{0}}(\bm{x}-\bm{a}_I)\psi_i(\bm{x})|^2,
     \quad\bm{n}\neq\bm{0},
     \end{equation}
   has the highest probability to yield the position signal.

   On the other hand, since energy preservation also follows from the exponential factor
   in the last line of Eq.~(\ref{eq:psif1tt}), we find for the elastically scattered piece
   of the wave function
   \begin{eqnarray}\nonumber
    \psi_f^{(\mathrm{el.})}(\bm{x})&=&\frac{m}{(2\pi)^3\mathrm{i}\hbar^2}\sum_{I=1}^N
    \int\!d^3\bm{y}\int\!d^3\bm{z}
  \int\!d^3\bm{x}'\int\!d^3\bm{k}\,\exp[\mathrm{i}\bm{k}\cdot(\bm{x}-\bm{z})]
  \\ \label{eq:psifel}
  &&\times
  \frac{\sin\!\left(k|\bm{z}-\bm{x}'|\right)}{
    \pi|\bm{z}-\bm{x}'|}\, V(\bm{z}-\bm{y})
  \phi^+_{\bm{0}}(\bm{y}-\bm{a}_I)\phi_{\bm{0}}(\bm{y}-\bm{a}_I)\psi_i(\bm{x}').
   \end{eqnarray}

 Except for extremely low-energy electrons, the remaining sinc function still has at most 
 {\AA}ngstr\"om-scale width, and we can still approximate short-range factors
 with $\delta$-functions,
   \begin{eqnarray}
     \psi_f^{(\mathrm{el.})}(\bm{x})&\propto&\sum_{I=1}^N
     \phi^+_{\bm{0}}(\bm{x}-\bm{a}_I)\phi_{\bm{0}}(\bm{x}-\bm{a}_I)\psi_i(\bm{x}).
   \end{eqnarray}
This would cut $N$ pieces of widths of order $\sqrt{\hbar/M\Omega}$ out of the incoming
wave function.
However, this time we cannot expect a quantum jump as a consequence of energy transfer
and the superposition principle persists.
Elastic scattering off the oscillator array therefore produces electron diffraction,
and the coherent superposition
of scattered waves from the $N$ scattering centers will imprint an interference
pattern on the elastically scattered wave function. Follow-up with inelastic scattering
in a second oscillator array for position detection, and accumulation of many single-particle
signals, will reveal that interference pattern.

\section{Conclusions}\label{sec:conc}

Heisenberg has suggested that continuous wave function evolution represents only a continuous
evolution of probability amplitudes for quantum jumps if energy transfer is involved.
On the other hand,
observation requires energy transfer between the observed system and the detector,
and quantum jumps would break continuous wave function evolution, thereby also
breaking the superposition principle. However, breaking the superposition principle
of wave function evolution
is exactly what is needed to avoid prediction of mutually contradictory pointer states
in low-energy quantum mechanics.
Mutually contradictory pointer states would correspond, e.g.,
to macroscopically separated position signals in single-particle
experiments.

The picture that emerges from this investigation favors an interpretation of the wave function
as an epistemic superposition of possible ontic outcomes, where elastic scattering corresponds to one
possible outcome, while every inelastic scattering channel corresponds to another
possible outcome.
Interference effects from elastic scattering then accounts for wavelike behavior, while
inelastic scattering yields particle-like behavior in observations.

The time-dependent Schr\"odinger equation does not evolve a quantum system \textit{per se},
but it only evolves relative probability amplitudes for different channels of possible
ontic outcomes of evolution of the system. The different channels are labelled by quantum
jumps related to energy transfer within a multi-partite system (or between systems),
or absence thereof. By the same token, the time-dependent Schr\"odinger equation
describes evolution of our best possible knowledge about evolution of a quantum system,
until observation updates our knowledge.

%  \section*{Data availability statement:}
  
%  No data were generated or analysed during this study. 

\acknowledgments
We acknowledge support from the Natural Sciences and Engineering Research Council
of Canada.\\

\end{document}